# Approaching ideal visibility in singlet-triplet qubit operations using energy selective tunneling-based Hamiltonian estimation

Jehyun Kim[1†], Jonginn Yun[1†], Wonjin Jang[1†], Hyeongyu Jang[1], Jaemin Park[1], Younguk Song[1], Min-Kyun Cho[1], Sangwoo Sim[1], Hanseo Sohn[1], Hwanchul Jung[2], Vladimir Umansky[3], and Dohun Kim[1]*

[1]*Department of Physics and Astronomy, and Institute of Applied Physics, Seoul National University, Seoul 08826, Korea*

[2] *Department of Physics, Pusan National University, Busan 46241, Korea*

[3]*Braun Center for Submicron Research, Department of Condensed Matter Physics, Weizmann Institute of Science, Rehovot 76100, Israel*

[†]*These authors contributed equally to this work.*

*Corresponding author:* dohunkim@snu.ac.kr

## Abstract

We report energy selective tunneling readout-based Hamiltonian parameter estimation of a two-electron spin qubit in a GaAs quantum dot array. Optimization of readout fidelity enables a single-shot measurement time of 16 μs on average, with adaptive initialization and efficient qubit frequency estimation based on real-time Bayesian inference. For qubit operation in a frequency heralded mode, we observe a 40-fold increase in coherence time without resorting to dynamic nuclear polarization. We also demonstrate active frequency feedback with quantum oscillation visibility, single-shot measurement fidelity, and gate fidelity of 97.7%, 99%, and 99.6%, respectively, showcasing the improvements in the overall capabilities of GaAs-based spin qubits. By pushing the sensitivity of the energy selective tunneling-based spin to charge conversion to the limit, the technique is useful for advanced quantum control protocols such as error mitigation schemes, where fast qubit parameter calibration with a large signal-to-noise ratio is crucial.

The efficient and precise characterization of a quantum system is important for building scalable quantum technologies that are robust to noise stemming from a fluctuating environment [1,2]. Estimating Hamiltonian parameters faster than the characteristic noise fluctuation time scale is essential, where knowledge gained from the measurement is used for correcting control parameters [2-4]. Active measurement-based feedback for example is used to enhance quantum sensing [5,6]. For semiconductor quantum dot (QD)-based spin qubit platforms, Hamiltonian parameter estimation applied to GaAs has shown that the effect of quasi-static nuclear spin fluctuation can be strongly suppressed for both single spin [7] and singlet-triplet qubits [2]. While the development of spin qubits in nuclear noise-free group-IV materials such as $^{28}$Si shows impressive progress in increasing single spin qubit coherence times [8,9], two-qubit control fidelity is often impeded by charge noise, which is also often sufficiently non-Markovian [10] and hence suppressible. Thus, fast Hamiltonian learning methods are expected to be used for a wide range of materials in noisy intermediate-scale quantum systems.

The fast single-shot measurement of qubits with high fidelity is a prerequisite for enabling Hamiltonian estimation. Semiconductor spin qubit devices mostly utilize a nearby charge sensor, where spin states are distinguished via spin to charge conversion mechanisms such as energy selective tunneling (EST) [11,12] or Pauli spin blockade (PSB) [13]. While both mechanisms are applicable for the detection of single spin [11], singlet-triplet ($ST_0$) [13], and exchange only qubits [14], PSB-based readout has been predominantly used for real-time Hamiltonian estimation owing to its deterministic readout time and fast initialization capability [15]. However, direct application of PSB often suffers from small signal contrast due to sub-optimal sensor position relative to double quantum dot (DQD) or fast relaxation at the readout condition due to large magnetic field difference-induced singlet state tunneling or the effect of spin-orbit coupling [16]. Variants of PSB-based readout have been developed using electron latching mechanisms in sufficiently isolated quantum dots [17,18] or by mapping to states outside the qubit space [19] circumventing some of the PSB-readout's known disadvantages. For Si devices, high readout visibility has been demonstrated using both PSB and EST readout owing to relatively long relaxation time [20-22]. However, so far the experiments using GaAs devices showed visibility below 80% using PSB readout.

The EST-based single-shot readout, on the other hand, guarantees a signal contrast corresponding to a full electron charge and long relaxation time [23,24]. As the Hamiltonian learning efficiency is directly affected by the ideality of the likelihood function, the large signal-to-noise ratio (SNR) of the EST readout can potentially be used for real-time Hamiltonian parameter estimation. Because the EST readout suffers from the intrinsically probabilistic nature of electron tunneling, requiring a longer waiting time than the PSB readout [25], it is important to determine whether the current state-of-the-art sensitivity of the RF-charge sensor can provide an EST readout that is sufficiently fast and simultaneously has a large SNR to enable efficient qubit frequency estimation on the fly.

In this Letter, we demonstrate real-time Hamiltonian parameter estimation by EST-based single-shot readout with sub-MHz accuracy in qubit frequency verified by observing over a 40-fold increase in coherence time $T_2$* compared to that of bare evolution on the order of 20 ns in GaAs [13]. With frequency feedback, the single-qubit operation performance in terms of initialization, manipulation, and measurement fidelity is one of the best figures reported thus far for semiconductor spin qubits, providing a promising route for applying the EST-based single-shot readout method to various qubit operations.

The quantum system we study is an $ST_0$ qubit with a basis state singlet $|S\rangle$ and triplet-zero $|T_0\rangle$, formed by two gate-defined lateral QDs. Fig. 1(a) shows a scanning electron microscope image of a quantum dot device similar to the one we measured. Au/Ti gate electrodes were deposited on top of the GaAs/AlGaAs heterostructure, where a 2D electron gas is formed $70\,\mathrm{nm}$ below the surface. Focusing on the DQD denoted by green circles in Fig. 1(a), high-frequency voltage pulses combined with DC voltages through bias tees are input to gates $V_1$, $V_2$, and $V_M$. RF-reflectometry was performed by injecting a carrier frequency of $\approx 125$ MHz with an estimated power of -100 dBm at the Ohmic contacts and monitoring the reflected power through homodyne detection. The device was operated in a dilution refrigerator with base temperature $\approx 7$ mK and with an external magnetic field $H_{ext}$. The measured electron temperature is $\approx 72$ mK [26-28].

The qubit Hamiltonian is given by $H = \frac{J(\varepsilon)}{2}\sigma_z + \frac{\Delta B_z}{2}\sigma_x$, where $J(\varepsilon)$ is the exchange splitting between states $|S\rangle$ and $|T_0\rangle$ controlled by potential detuning $\varepsilon$, $\sigma_{i=x,y,z}$ is the Pauli matrix, and $\Delta B_Z$ is the magnetic field difference between QDs set by the hyperfine interaction with the host Ga and As nuclei. We adopted units where $g^* \mu_B / h = 1$, in which $g^* \approx -0.44$ is the effective gyromagnetic ratio in GaAs, $\mu_B$ is the Bohr magneton, and $h$ is Planck's constant. With the quantum control provided by rapidly turning on and off $J(\varepsilon)$, the main task is to estimate $\Delta B_z$, which varies randomly in time owing to statistical fluctuations of the nuclei. The basic idea of the Bayesian inference is to update one's knowledge about the Hamiltonian parameter by comparing the measurement results with the expected form of time evolution (likelihood function). Based on the single-shot projective measurement of the qubit evolving around the *x*-axis on the Bloch sphere for time $t_k = 4k$ ns (Larmor oscillation), Bayesian inference is performed by the following rule up to a normalization constant [2]:

$$P(\Delta B_z | m_N, m_{N-1}, \ldots m_1) = P_0(\Delta B_z) \prod_{k=1}^{N} \frac{1}{2}[1 + r_k(\alpha + \beta\cos(2\pi\Delta B_z t_k))] \quad (1)$$

where $N$ is the number of single-shot measurements per Hamiltonian estimation, $P_0(\Delta B_z)$ is the uniform initial distribution, $r_k$=1(-1) for $m_k = |S\rangle (|T_0\rangle)$, and $\alpha\,(\beta)$ is the parameter determined by the axis of rotation (oscillation visibility). After the $N^{\text{th}}$ single-shot measurement and update, the most probable $\Delta B_z$ is determined from the posterior distribution $P(\Delta B_z | m_N, m_{N-1}, \ldots m_1)$.

In the likelihood function $\frac{1}{2}[1 + r_k(\alpha + \beta\cos(2\pi\Delta B_z t_k)]$, ideally, $\alpha$=0 and $\beta$=1. Fig. 1(b) shows the simulation results of the root mean squared error between the true and estimated $\Delta B_z$. Compared to the low-visibility case ($\beta$= 0.5) corresponding to a large measurement error, the high-visibility case ($\beta$= 0.9) shows a large improvement in the rate of convergence, reaching sub-MHz accuracy in less than $N$ = 70. To date, Bayesian estimations of quantum dot spin qubits have been performed with $\beta$~0.7 [2,7] requiring $N$ > 120 for practical Hamiltonian estimation. Below, we show that the EST readout indeed provides $\beta$ reaching unity enabling efficient frequency detection and feedback.

Fig. 1(c) shows a schematic block diagram and an example scope trace during the experiment. We set the integration time of the RF demodulator $t_{\mathrm{int}} = 200$ ns, at which SNR = 9.2 [24,26,29-33]. The measurement time was set to 15 μs, during which the dot-to-reservoir tunnel rate tuned to the order of 1 MHz ensures that a tunnel-out event occurs for the state $|T_0\rangle$. For the probe sequence, we diabatically pulse $\varepsilon$ to rapidly turn off $J$. The calculation time according to Eq. (1) is $\approx 10\ \mu s$ after the $k^{\mathrm{th}}$ measurement. For the operation, there are two types of modes. The first is heralded mode where the operation is conditionally triggered only when the estimated qubit frequency in the probe step falls within a preset tolerance $\delta(\Delta B_{\mathrm{z}})_{\mathrm{set}}$ around the target frequency $\Delta B_{\mathrm{z.t}}$. Once a short operation on the order of 20 shots is finished, one has to wait for the next $\Delta B_{z,t} \pm \delta(\Delta B_{\mathrm{z}})_{\mathrm{set}}$ to happen. The method is conceptually similar to Ref. [34] where the Bayesian estimator-based heralding was used to effectively suppress thermally induced initialization error. The second is the active feedback mode where resonant modulation of $J(\varepsilon)$ (Rabi oscillation) is performed using the frequency obtained from the probe step. Here, $\delta(\Delta B_{\mathrm{z}})_{\mathrm{set}}$ is typically set to more than 70 MHz and the control frequency is actively adjusted so that the waiting time is minimized. In all steps, we apply an adaptive initialization step [34,35] where the controller triggers the next experiment provided that the state is $|S\rangle$. Including all the latency components, the repetition period for one probe (operation) step is approximately 26 (16) μs on average [26]. Fig. 1(d) shows typical histograms of $\Delta B_{\mathrm{z}}$ obtained by repeatedly running the probe step at different $H_{\mathrm{ext}}$, showing fluctuation about a non-zero mean $\Delta B_{\mathrm{z}}$. Note that the average $\Delta B_z$ depends on $H_{ext}$. While the exact origin of this is not well understood to date, previous studies in GaAs quantum dot report similar behavior [36,37], and we adjust $H_{ext}$ to set the most probable $\Delta B_z$ about 30 MHz (110 MHz) for the heralded (active feedback) mode.

First, we demonstrate the performance of the EST-based Bayesian estimator using the heralded mode operation. Fig. 2(a) shows the representative Larmor oscillations where $P_1$ is the triplet return probability with $N$ = 70, $\Delta B_{z.t}$ = 30 MHz, and $\delta(\Delta B_z)_{set}$ = 0.1 MHz. The measurement of $T_2$*($N$), extracted by fitting the Larmor oscillations to a Gaussian decay, reveals the uncertainty of the EST-Bayesian estimation (Fig. 2(b)). The initial increase in $T_2$*($N$) corresponds to an improvement in the estimation accuracy. $T_2$* reaches an optimal coherence time of over 800 ns near $N$ = 70 and subsequently decreases for $N$ > 80. The latter reflects the effect of nuclear fluctuation during the increased estimation period consistent with the diffusive behavior of $\Delta B_z$ with diffusivity D = 10.16 kHz$^2$/μs (Fig. 2(c)) [2].

Fig. 2(d) shows the $\delta(\Delta B_z)_{set}$ dependence of the experimental estimation uncertainty $\sigma_{\Delta B_z} = 1/\sqrt{2}\pi T_2*$ [38]. As we set the tolerance more stringently (smaller $\delta(\Delta B_z)_{set}$), $T_2$* increases correspondingly. The residual uncertainty of the EST-based Bayesian estimator when $\delta(\Delta B_z)_{set}$ = 0 is approximately 0.25 MHz. It is likely overestimated by the nuclear fluctuation during the operation time of 0.32 ms (16 μs × 20 shots) after the probe step. Thus, we conclude that our Hamiltonian estimation scheme enables qubit frequency estimation in 70 shots with an accuracy better than 0.25 MHz. Note also that while the maximum $T_2$* = 835 ns we observe is less than the PSB-based Hamiltonian estimation [2], the actual performance of the PSB and EST-based Bayesian estimators is difficult to directly compare so far because the dynamic nuclear polarization [3,39] is not used in the current experiment.

We now discuss the application of the EST-based Hamiltonian estimation to general single-qubit operations (Fig. 3: heralded mode, Fig. 4: active feedback mode). Fig. 3(a) shows coherent Larmor oscillations with $\Delta B_{z,t}$ = 30 MHz. The oscillation shows the visibility of approximately 97.7%. Considering possible imperfections in the control stemming from residual J and finite rise time of the waveform generator (~0.4 ns), the result shows that the EST-based Bayesian method enables accurate qubit frequency estimation and high measurement fidelity at the same time, leading to near ideal visibility. By comparing the oscillation with the numerical simulation, we estimate measurement fidelity of 99% with less than 0.1% initialization errors for the heralded mode [23,26,40,41].

Using symmetric barrier-pulse operation, recently demonstrated in Ref. [42], Fig. 3(b) shows coherent exchange oscillations with $\Delta B_{z,t}$ = 30 MHz, and $J$ = 75 MHz. In addition, a two-dimensional map of the exchange oscillations is measured as a function of exchange amplitude $A_{ex}$ and exchange duration $t_e$ (Fig. 3(c)), showing the oscillations with a high-quality factor $Q$. Moreover, $Q(J)$ follows the general trend observed in previous results [42] where $Q$ ($T_{decay}$) tends to saturate (decrease) at large $J$ owing to the crossover from nuclear noise to electrical noise-limited decoherence. While the maximum $Q$ of ~40 is comparable to that in the previous report [42], our EST-based Bayesian method effectively suppresses the $\Delta B_z$ fluctuation, leading to the observation of Q > 30 in a wide range of $J$.

Although the heralded mode operation exemplifies the performance of the EST-based Hamiltonian estimator with minimal overhead in the Bayesian circuit, the main drawback is the low duty cycle (actual operation/waiting time), which can be < 1% depending on the tolerance. Thus we further develop our methodology using ac-driven qubit operation in active feedback mode. The pulse sequence for qubit operation is the same as in Fig. 3(b) except that a sinusoidal RF pulse is applied to $V_M$ using the frequency detected in the probe step. In this manner, the total waiting time is reduced down to one probe step (70 shots x 26 μs = 1.82 ms). Fig. 4(a) shows the coherent Rabi oscillation measured as a function of the RF pulse duration and controlled detuning $\delta f$. The pulse amplitude $A_{RF}$ is chosen to maximize the $Q$ factor $Q_{Rabi} = f_{Rabi}T_{Rabi} \approx 12$ with the Rabi frequency $f_{Rabi}$ of 6.05 MHz and the Rabi decay time $T_{Rabi}$ of 1.71 μs (inset to Fig.4 (a)). The oscillation visibility reaches approximately 97.6 %, (Fig. 4(b)). This near-ideal visibility of the RF-driven oscillation even without dynamic nuclear polarization again reveals the precise qubit frequency estimation and high measurement fidelity simultaneously enabled by the EST-based Bayesian estimator.

Furthermore, we perform the standard randomized benchmarking (RB) and interleaved randomized benchmarking (IRB) where single-qubit gates X, Y, X/2, Y/2, −X/2, and −Y/2 are interleaved to random Clifford gates [43-45]. The recovery gate is chosen such that the final state is ideally singlet, and the gate fidelity is obtained by fitting the measured data to the exponentially decaying curve [26,45]. We find the average gate fidelity $F_{avg}$ of 96.80 % and $\pi$-pulse fidelity $F_X$ of 99.13 %, the latter being close to the $Q$-factor limited value $e^{-1/(2Q_{Rabi})^2} = 99.76 \pm 0.03$ %.

To compare the state preparation and measurement (SPAM) errors between two operation modes, we perform gate-set tomography (GST) [41]. Fig. 4(d) shows the density matrix (top row) and the Pauli transfer matrix (PTM, bottom row), obtained using a single qubit GST protocol with a gate set {I, X/2, and Y/2}2 [26,46], from which we obtain $F_{X/2}$ = 99.05 % and $F_{Y/2}$ =98.2 %, consistent with the values obtained from the IRB. The GST yields the initialization fidelity of 99.7% and measurement fidelity of 98.3%. We ascribe slightly lower initialization and measurement fidelity for the active feedback mode-based GST compared to the heralded mode to a combination of an additional leakage probability through S-$T_+$ anticrossing while preparing (projecting) a state on the $x(z)$-axis of the Bloch sphere and the increased relaxation probability during the idle time between the discrete gates. Nevertheless, these results consolidate the high gate fidelity and low SPAM error illustrating that our Hamiltonian estimation enables the real-time application of general qubit operations in GaAs with the fidelities reaching the level of singlet-triplet qubits in Si devices [47].

In conclusion, using energy selective tunneling readout-based Hamiltonian parameter estimation of an $ST_0$ qubit in GaAs, we demonstrated passive and active suppression of nuclear noise, leading to $T_2$* above 800 ns, near-ideal quantum oscillation visibility, and gate fidelity up to 99.6% confirmed by both RB and GST comparable with recently demonstrated optimal control-based gate fidelity [48]. The work showcases the improvements in the overall capabilities of GaAs-based spin qubits. With the large SNR of the charge sensor and real-time capability, the EST-based Hamiltonian estimation is potentially useful for advanced quantum control protocols with affordable overhead in classical signal processing, such as error mitigation schemes and entanglement demonstration experiments, where fast qubit parameter calibration with large readout visibility is essential [35].

**Acknowledgments**

This work was supported by the Samsung Science and Technology Foundation under Project Number SSTF-BA1502-03. The cryogenic measurement used equipment supported by the National Research Foundation of Korea (NRF) Grant funded by the Korean Government (MSIT) (No. 2018R1A2A3075438, No.2019M3E4A1080144, No.2019M3E4A1080145, and No.2019R1A5A1027055), Korea Basic Science Institute (National Research Facilities and Equipment Center) grant funded by the Ministry of Education (No.2021R1A6C101B418), and the Creative-Pioneering Researchers Program through Seoul National University (SNU). Correspondence and requests for materials should be addressed to D.K. (dohunkim@snu.ac.kr )

**Figure captions**

**Fig. 1. (a)** Scanning electron microscopy image of a device similar to the one used in the experiment. Green (yellow) circles indicate the position of quantum dots for the $ST_0$ qubit (RF charge sensor. $H_{ext}$ is applied to the $z$-axis as indicated by the blue arrow. **(b)** Root mean squared error of the Bayesian estimator as a function of $N$ and $\beta$. **(c)** Left panel: block diagram of the experimental procedure including the probe and operation step, where the latter is performed either in heralded or active feedback mode. Right panel: example scope trace of the charge sensor signal recorded during the experiment. Gray trace: RF-demodulated sensor signal with SNR = 9.2 at $t_{int}$ =200 ns. Blue trace: trigger signals marking the start timings of each probe and operation step. The red dots show the timings of the initialization check sequences. **(d)** Histograms of $\Delta B_z$ obtained by running the probe step 10000 times at two different $H_{ext}$. For the heralded (active feedback) mode, $\delta(\Delta B_z)_{set}$ on the order of 1 MHz (few tens of MHz) around an average $\Delta B_z$ of 30 (110) MHz was chosen. Green dashed lines indicate a tolerance window $2\delta(\Delta B_z)_{set}$ .

**Fig. 2. (a)** Representative Larmor oscillations with $N$ = 70 showing $T_2$* = 835 ns, with a fit to a Gaussian decay function (red envelope and blue oscillatory fit). **(b)** $T_2$* as a function of $N$, showing an optimal $N$ = 70 with $\delta(\Delta B_z)_{set}$ = 0.1 MHz. **(c)** The variance of the $\Delta B_z$ as a function of elapsed time showing a diffusion process with the diffusivity $(10.16 \pm 0.06\,\mathrm{kHz})^2 / \mu\mathrm{s}$ . **(d)** The uncertainty of the frequency estimation $\sigma_{\Delta B_z}$ as a function of the half-width of the tolerance $\delta(\Delta B_z)_{\mathrm{set}}$ .

**Fig. 3.** **(a)** Top: Pulse sequences applied to gates $V_1$ and $V_2$ for the heralded Larmor oscillations measurement. Bottom: Larmor oscillations with visibility higher than 97% **(b)** Top: Pulse sequence for coherent exchange operation. Bottom: Corresponding exchange oscillations at $J$ = 75 MHz, $\Delta B_{z,t}$ = 30 MHz showing charge noise-limited coherence time $T_{\mathrm{decay}}$ = 450 ns. **(c)** Exchange oscillations as a function of barrier pulse amplitude $A_{\mathrm{ex}}$ and evolution time $t_e$. **(d)** $T_{\mathrm{decay}}$ and the quality factor $Q$ as a function of exchange coupling $J$.

**Fig. 4.** **(a)** Rabi oscillation of $P_1$ as a function of controlled detuning $\delta f$ and pulse duration. Inset: Oscillation quality factor $Q_{\mathrm{Rabi}}$ as a function of RF amplitude $A_{\mathrm{RF}}$ (measured at the output of the signal generator). The red symbol marks the condition for the maximum $Q_{\mathrm{Rabi}}$. **(b)** Representative Rabi oscillation with visibility higher than 97 %. The oscillation is fit to the sinusoidal function with the Gaussian envelope, from which Rabi decay time $T_{\mathrm{Rabi}}$ = 1.71 $\mu s$ is obtained. **(c)** $P_1$ as a function of the number of random Clifford gates obtained from a single qubit standard and interleaved randomized benchmarking. Traces are offset by 0.3 for clarity. **(d)** Density matrices (top row) and Pauli transfer matrices (bottom row) evaluated by gate set tomography.

## Figures

Figure 1

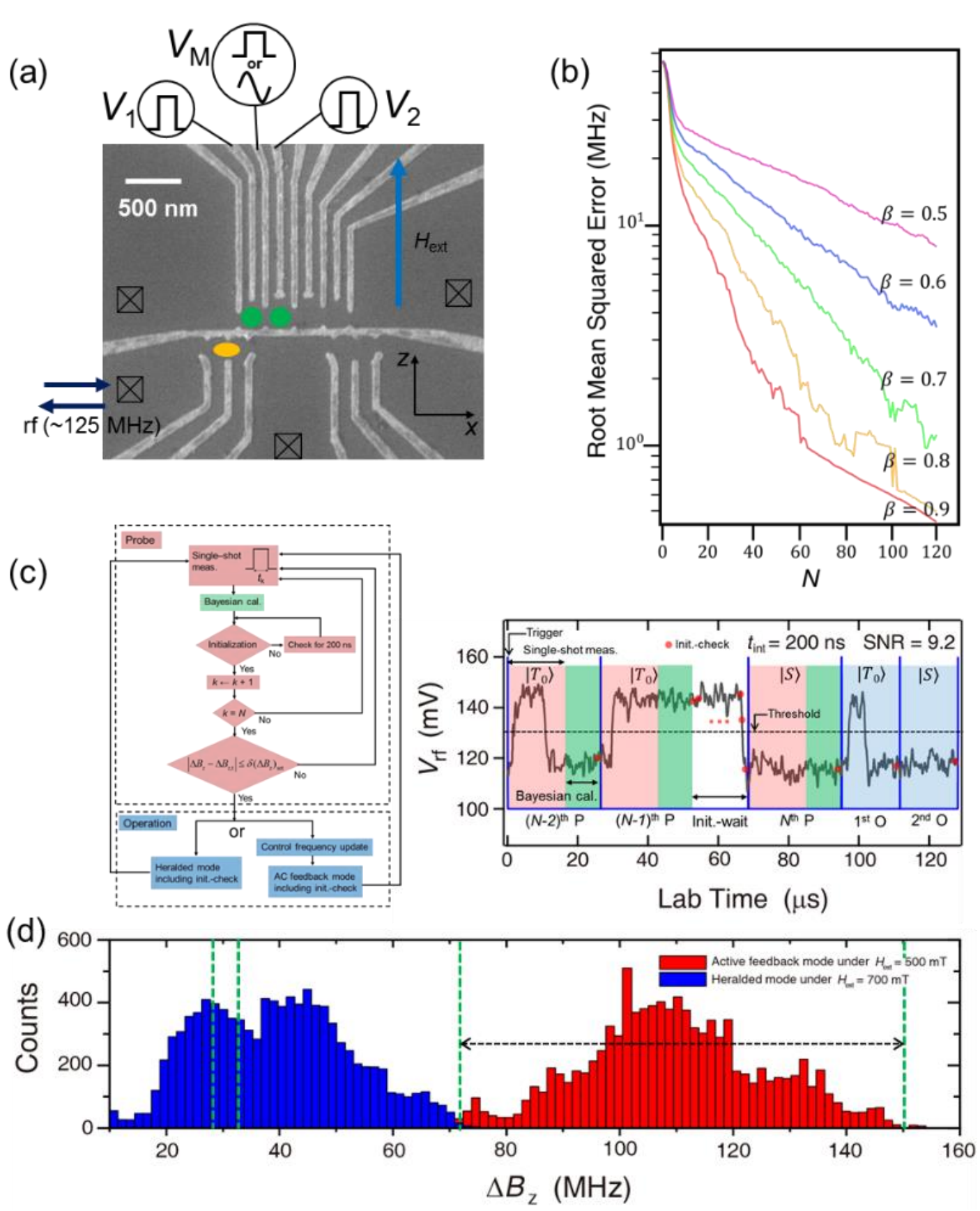

Figure 2

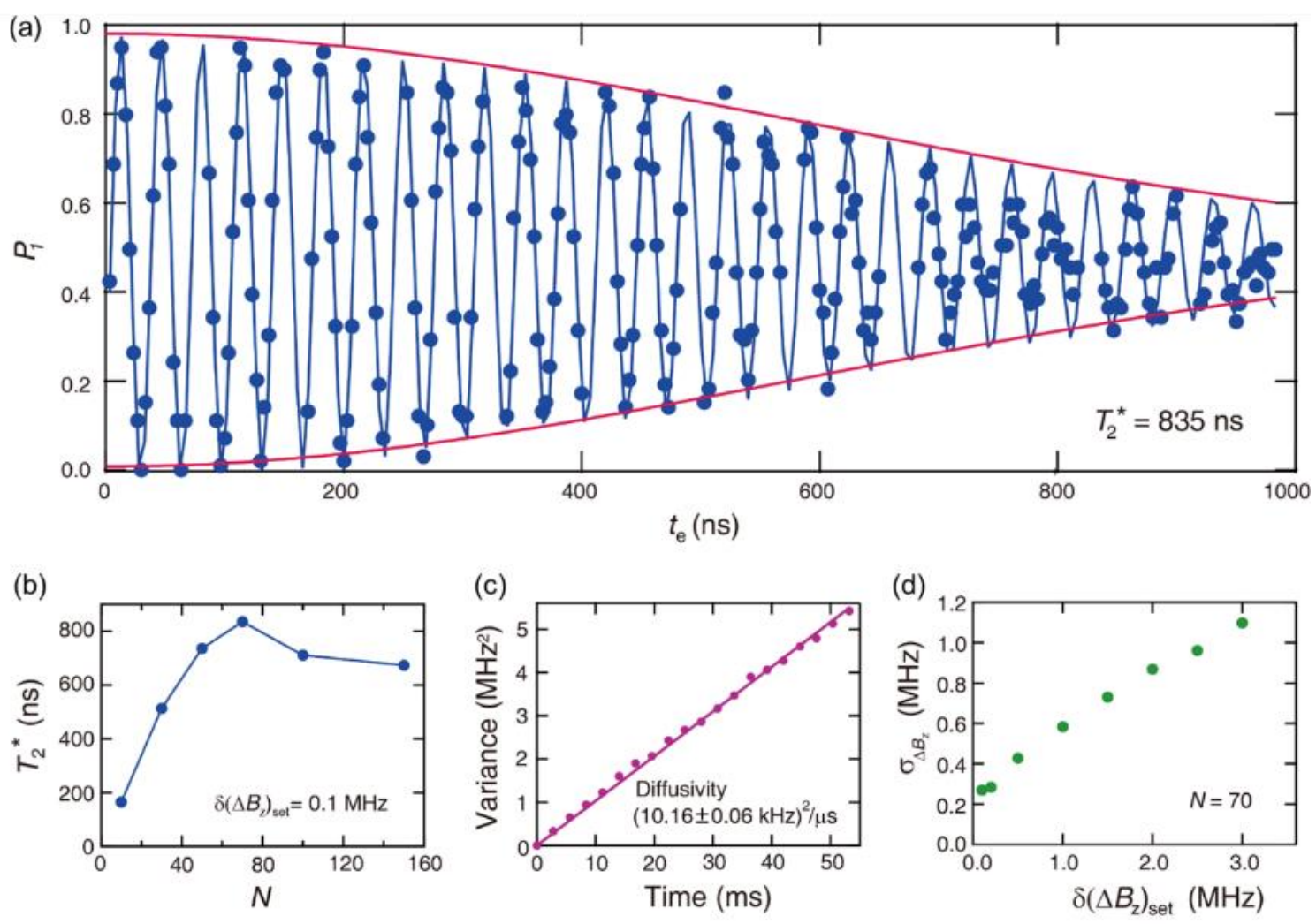

Figure 3

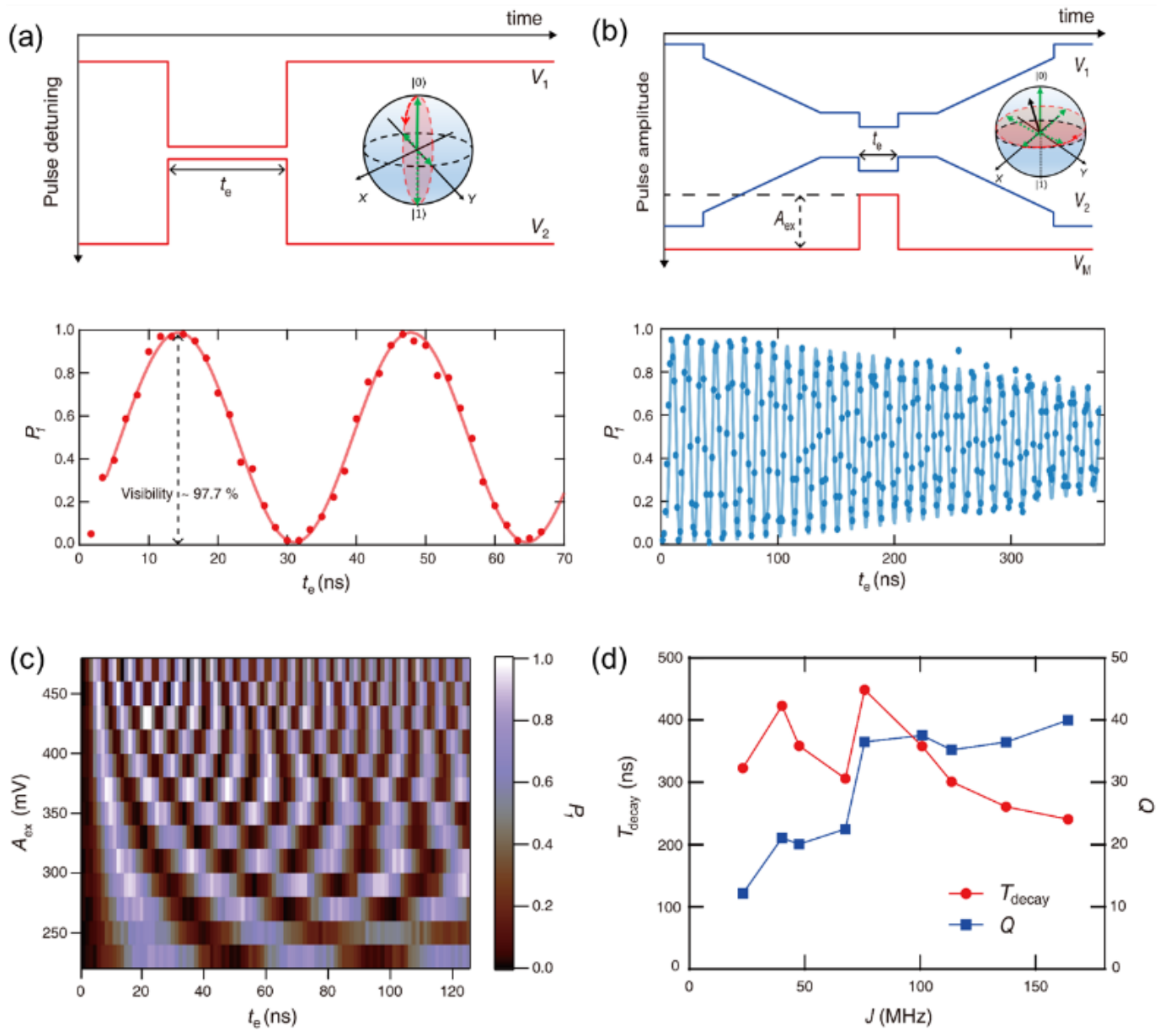

Figure 4

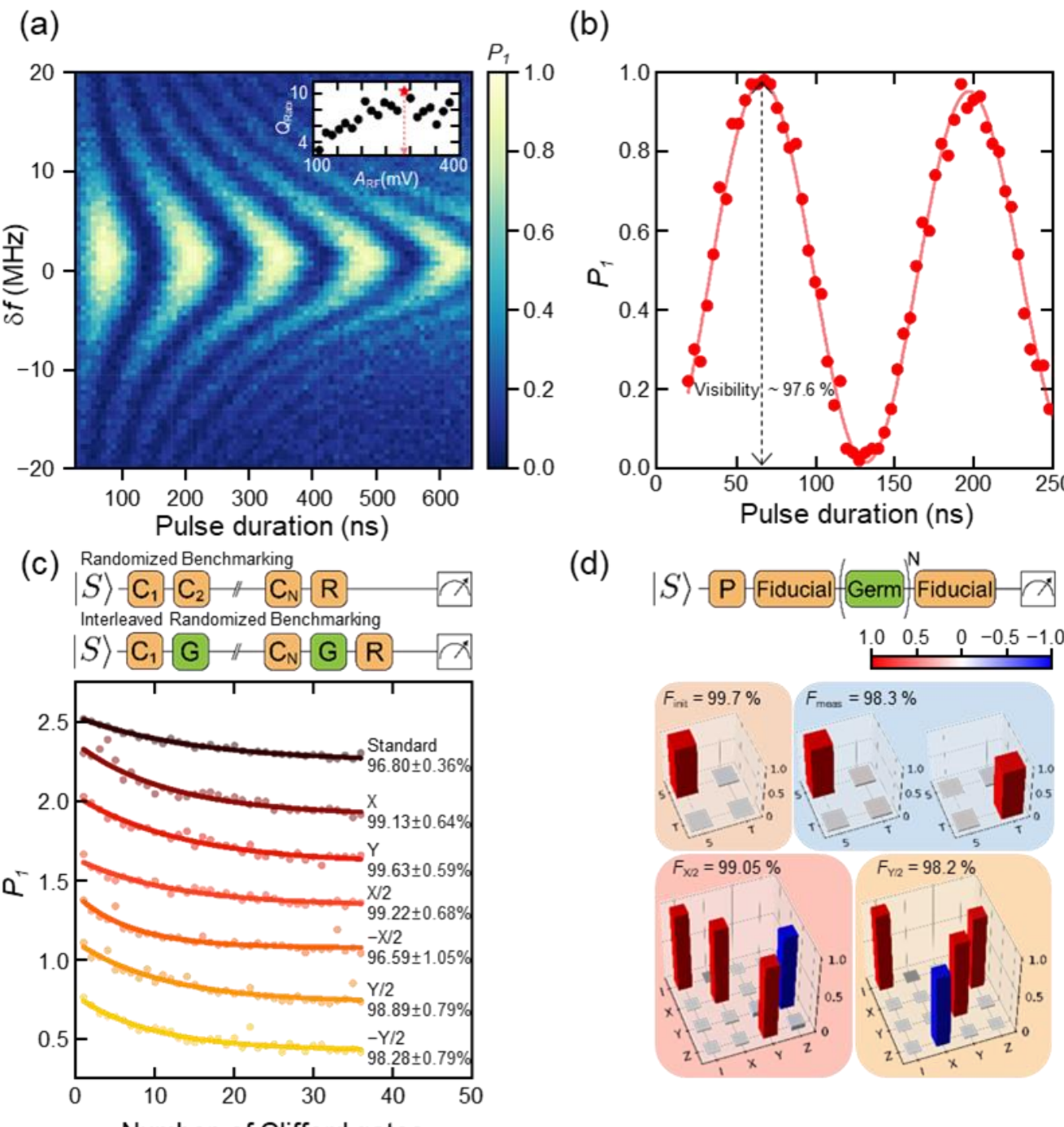

**Supplementary Materials**

## S1. Measurement setup and FPGA implementation

An RF-single electron transistor (RF-SET) sensor is used to detect the quantum states of the $ST_0$ qubit. An impedance matching tank circuit as shown in Fig. S1 is attached to the RF-ohmic contact of the device. With the inductor value $L$ = 1500 nH and the parasitic capacitance $C_p$ = 1.4 pF of the circuit board, the resonance frequency is about 125 MHz, and the impedance matching occurs when the conductance of the RF-SET sensor is approximately 0.5 $h/e^2$ where $h$ is Plank's constant and $e$ is the electron charge. A commercial high-frequency lock-in amplifier (Zurich Instrument, UHFLI) is used as the carrier generator, RF-demodulator for the homodyne detection, and further signal processing units such as gated integration and timing marker generation. Carrier power of – 40 dBm is generated at room temperature and further attenuated through the cryogenic attenuators and the directional coupler by -60 dB. The reflected signal is first amplified by 50 dB with a two-stage commercial cryogenic amplifier (Caltech Microwave Research Group, CITLF2 x 2 in series), and further amplified by 25 dB at room temperature using a home-made low-noise RF amplifier.

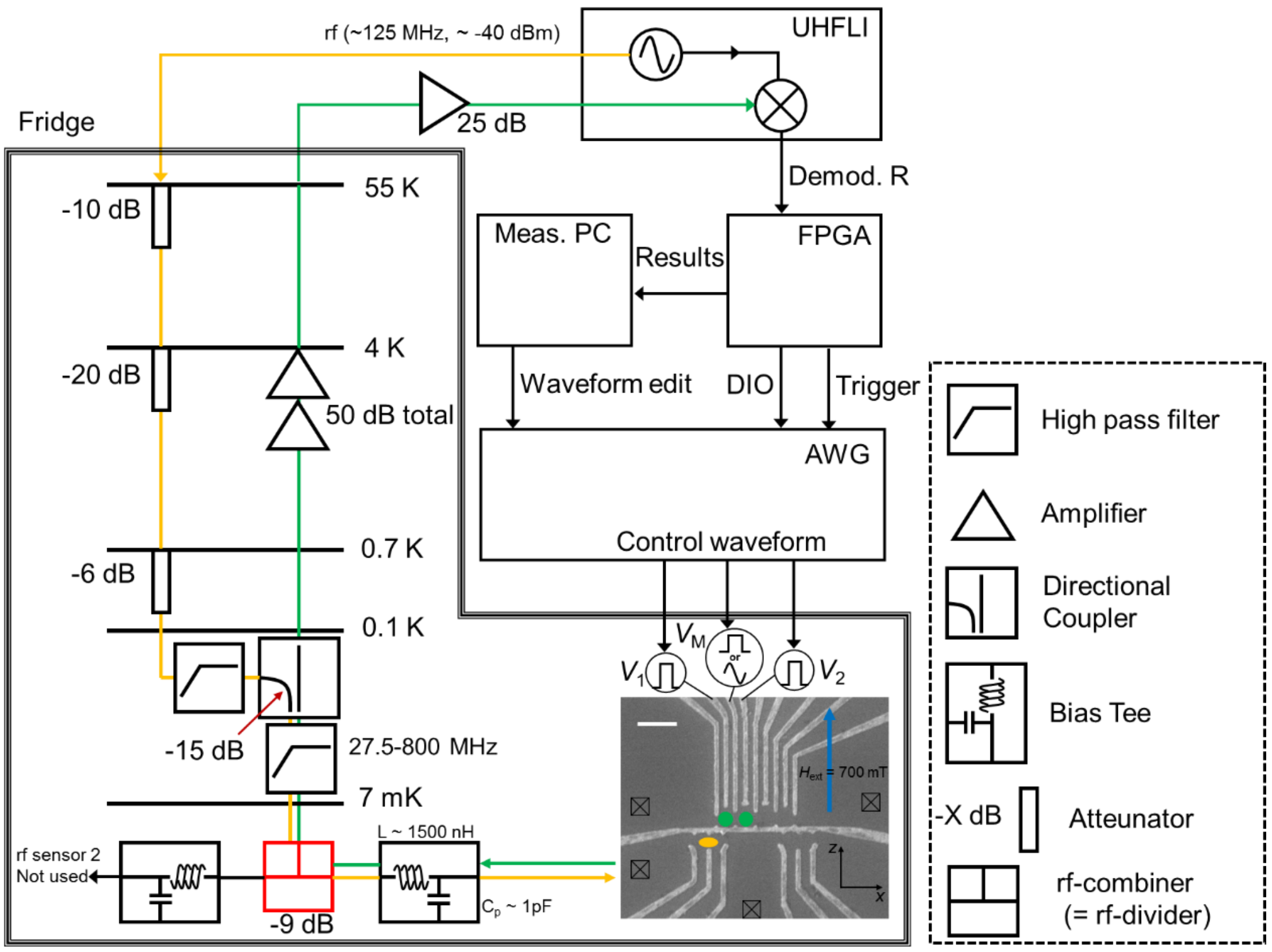

**Supplementary Figure S1.** Measurement setup for radio frequency (RF)-reflectometry and signal block diagram. An impedance matching tank-circuit (L ~1500 nH, $C_p$ ~1.4 pF) is attached to the RF-SET sensor Ohmic contact for homodyne detection. The yellow (green) line indicates the input (reflected) signal. The reflected signal is demodulated in the UHFLI, and subsequently processed in a Field Programmable Gate Array (FPGA) for the EST readout-based Bayesian estimation.

For real-time data processing, we implement a digital logic circuit with a Field Programmable Gate Array board (FPGA, Digilent Zedboard with Zynq-7000 XC7Z020-CLG484). The RF-demodulated analog signal from the UHFLI is input to the 12-bit ad-converter of the FPGA. For single-shot discrimination, the transient tunneling events of the qubit state are thresholded in real-time by comparing the preset threshold value with the data in parallel. The discriminator records bit 1 immediately when data above the threshold value is detected. The bit 0 is recorded when such events did not happen throughout the preset measurement period of 15 μs. The Bayesian estimation after a single shot measurement for the probe step is carried out by calculating the posterior probability distribution for 512 values of $\Delta B_z$ between 10 and 160 MHz. We use a look-up table (LUT) storing all the possible values of the likelihood function in the Block RAM inside the FPGA and design a 512-parallelized calculation module to minimize latency due to data processing. After the calculation, the FPGA follows either of the following steps depending on the operation mode. For the heralded mode operation, the user-defined controller triggers the operation step provided that the $\Delta B_z$ calculated after the $N^{\text{th}}$ Bayesian update is in the range $\Delta B_{z,t} \pm \delta(\Delta B_z)_{\text{set}}$ where $\Delta B_{z,t}$ is the target frequency and $\delta(\Delta B_z)_{\text{set}}$ is the preset tolerance. For the active feedback mode, the FPGA converts the estimated $\Delta B_z$ into a 9-bit digital signal and sends it to the digital input/output port of the arbitrary waveform generator (Zurich Instruments, HDAWG). The HDAWG applies the square-wave enveloped sinusoidal waveform with the frequency corresponding to the digital value to $V_M$ using the multifrequency modulation function. For both probe and operation steps, an adaptive state initialization is performed by acquiring a 200 ns long sample and thresholding repeatedly until the lastest value falls below the threshold. For the entire data processing, about 60% of LUT and 38% of Flip Flop resources were used.

## S2. Charge stability diagram and electron temperature

Fig. S2(a) shows the charge stability diagram as a function of gate voltages $V_1$ and $V_2$ showing the relevant region for the EST-Bayesian of our $ST_0$ qubit, where initialization/read-out points in (2,0) and the operation point in (1,1) are depicted as black circles. Fig. S2(b) shows the normalized charge transition signal of the last electron in the left quantum dot as a function of $V_1$ at the mixing chamber temperature $T_{\text{mixing}}$ = 7 mK. This data is fitted to the Fermi-Dirac distribution curve given by $P_e(V_1) = \frac{1}{e^{a(V_1 - \text{b})} + 1}$, $a = \frac{\alpha}{k_B T_e}$, where $a$ and $b$ are fitting parameters, $\alpha$ is the lever-arm for $V_1$, $k_B$ is the Boltzmann constant, and $T_e$ is the electron temperature. The $1/a$ extracted at several different $T_{\text{mixing}}$ is converted to $T_e$ using $\alpha$ = 0.0497 meV/mV obtain from the linear relationship for $T_{\text{mixing}}$ > 100 mK as shown in Fig. S2(c) [1]. From a power law $T_e(T_{\text{mixing}}) = (T_S^k + T_{\text{mixing}}^k)^{\frac{1}{k}}$ where $T_S$ is a saturation limit of $T_e$ at $T_{\text{mixing}}$ = 0 mK and $k$ is an exponent that depends on the thermalization mechanisms, we estimate $T_S$ = 72 mK and $k$ = 3.35, indicating that Wiedemann-Franz cooling is a dominant cooling mechanism rather than electron-phonon cooling [2].

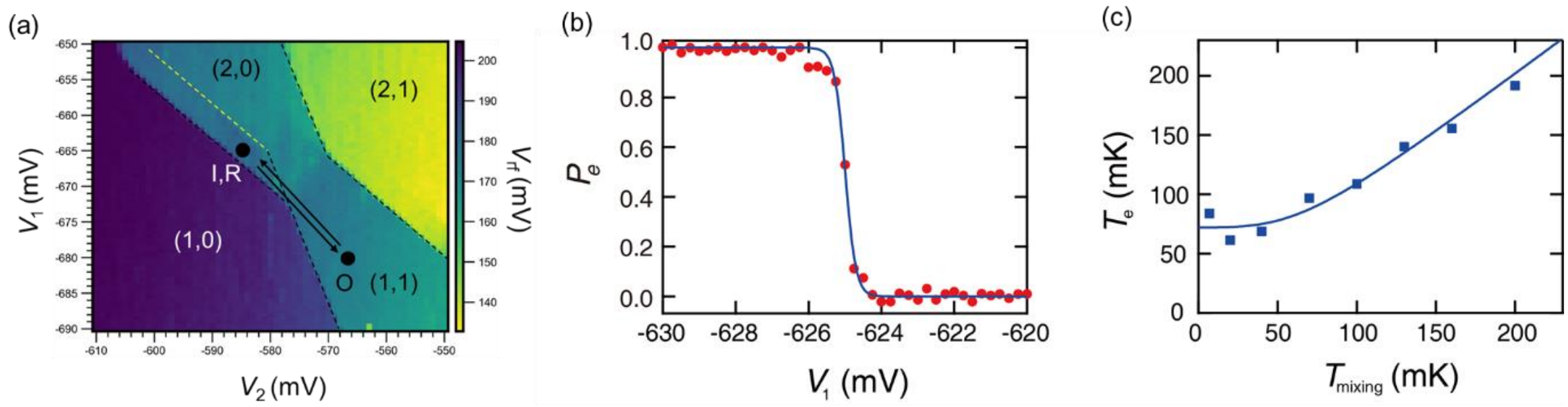

**Supplementary Figure S2. (a)** Charge stability diagram measured at the mixing chamber temperature $T_{\text{mixing}}$ = 7 mK. The Yellow dashed line indicates the boundary of the EST-readout window. **(b)** Normalized charge transition signal from (1,0) to (0,0) as a function of $V_1$ at $T_{\text{mixing}}$ = 7 mK. **(c)** Electron temperature $T_e$ extracted from broadening of the Fermi-Dirac distribution as a function of $T_{\text{mixing}}$ showing estimated $T_e$ of 72 mK at $T_{\text{mixing}}$ = 7 mK.

### S3. Charge sensitivity

We evaluate the sensitivity of the charge sensor by observing the integration time $t_{\text{int}}$ dependence of the signal-to-noise ratio (SNR). We define the SNR by $\Delta V / \sigma$, where $\Delta V$ is the sensor signal contrast for a single electron charge transition and $\sigma$ is the rms noise amplitude at a given $t_{\text{int}}$. The sampling rate of the oscilloscope is set above 200 MHz. As shown in Fig. S3, the SNR is proportional to $\sqrt{t_{\text{int}}}$ and we linearly fit the $\text{SNR}^2$ to extract the minimum integration time for achieving SNR = 1, $\tau_{\min}$ of 2.45 ns [3].

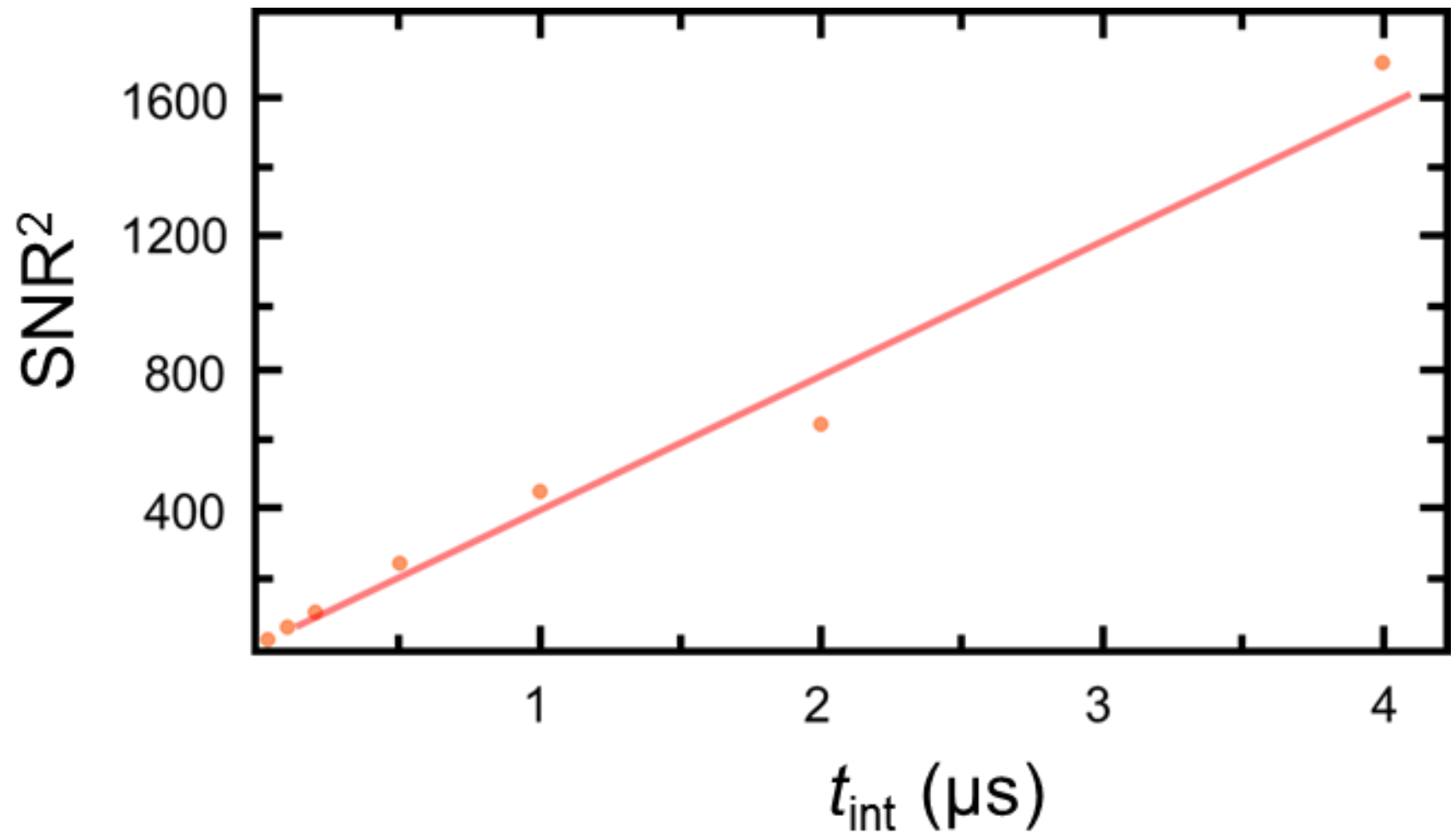

**Supplementary Figure S3.** Signal to noise ratio (SNR) of the RF-single-electron transistor charge sensor as a function of integration time $t_{int}$. The minimum integration time $\tau_{min} \sim 2.45$ ns corresponding to the integration time for achieving the unit SNR is obtained from extrapolating a linear fit to the data.

Using $\tau_{min}$ as a suitable metric for binary charge detection sensitivity $e\sqrt{\tau_{min}}$ [4,5], we compare performances of the recently published works as shown in Supplementary Table 1. [3,4,6-8] showing that the charge sensitivity achieved in this work is one of the best values available. By comparison, the charge sensor used in this work is more sensitive than a dispersive sensor with a cavity-coupled Josephson parametric amplifier [3] but less sensitive than a similarly prepared RF-SET in a strong quantum dot – sensor capacitive coupling regime [4].

| Literature | $\tau_{min}$ (ns) | Charge Sensitivity ($e\sqrt{\mathrm{Hz}}$) |
|---|---|---|
| C. Barthel *et al.* (2009)[Ref. 5] | 400 | $6.32 \times 10^{-4}$ |
| C. Barthel *et al.* (2010)[Ref. 6] | 23 | $1.52 \times 10^{-4}$ |
| J. Stehlik *et al.* (2015)[Ref. 2] | 7 | $8.37 \times 10^{-5}$ |
| D. Keith *et al.* (2019)[Ref. 3] | 1.25 | $3.54 \times 10^{-5}$ |
| A. Noiri *et al.* (2020)[Ref. 7] | 38 | $1.95 \times 10^{-4}$ |
| Our work (2022) | 2.45 | $4.95 \times 10^{-5}$ |

**Supplementary Table 1.** Comparison of minimum integration time $\tau_{\min}$ and corresponding charge sensitivity.

## S4. Visibility analysis

We analyze the visibility of the quantum oscillation shown in Fig. 3 in the main text with a numerical model which includes the thermal tunneling, and the false initialization errors. The analysis essentially amounts to combining the visibility with the computed readout infidelities to extract the relevance of other effects. We first evaluate the tunneling detection infidelity of our readout circuit by numerically simulating the histogram of the RF single-shot traces [9,10]. Following the Ref. 10, we fit the numerical histogram obtained from the simulated traces to the experimental histogram which yields the tunneling detection error (Fig. S4(a)) of $E_T$ ($E_N$) ~ 1.4 % (0.7 %) where the $E_T$ ($E_N$) corresponds to the infidelity for detecting the tunneling (no-tunneling) events.

Based on the tunneling detection infidelities, we extract the state measurement fidelities by fitting the Larmor oscillation curve to the numerical model which comprises the state relaxation, false initialization, and the thermal tunneling errors, where the following parameters describe the error rates respectively.

$\alpha_S$ : Thermal tunneling probability of the singlet ($S$) state

$\beta_{T(S)}$ : Probability for the qubit state to be initialized to the triplet (singlet) state

: Relaxation probability ~ $\tau_{out}/T_1$ ~ 0.3% where we use $T_1$ ~ 337 μs previously measured in Ref. 10 as a rough estimate. While $T_1$ time can be different depending on tuning conditions, we obtain measurement fidelity consistent with that of gate set tomography (see section S5 below).

With $P_{flip}(\tau)$ ~ $\sin^2(\pi\Delta B_z\tau)$ corresponding to the ideal diabatic Larmor oscillation under the magnetic field gradient $\Delta B_z$, we estimate the probability $P_i(\tau)$ (i = $S$, $T_0$, $T_+$, $T_-$), which is the realistic probability for the qubit state to be at one of the two-spin states after the manipulation. We assume the polarized triplet states $T_+$, and $T_-$ states are not involved in the coherent dynamics at the manipulation stage, and all three triplet states have the same relaxation rates to the ground (singlet) state. We also suppose that false initialization probability to each of three triplet states is all equal to $\beta_T/3$. The estimation procedure is as follows.

i) $P_S(\tau)$ : Probability for the final qubit state to be $S$ after the manipulation.

- Initializes to $S$ ($\beta_S$), does not flip under the manipulation pulse (1- $P_{flip}(\tau)$)
- Initializes to $S$ ($\beta_S$), flip under the manipulation pulse ($P_{flip}(\tau)$), relax to the ground state ($\gamma$)
- Initializes to $T_0$ ($\beta_T/3$), flip under the manipulation pulse ($P_{flip}(\tau)$)
- Initializes to $T+$ or $T-$ ($2\beta_T/3$), relax to the ground state ($\gamma$)

$$\Rightarrow P_S(\tau) = \beta_S[1-P_{flip}(\tau)+P_{flip}(\tau)\gamma] + {}^{\beta_T}\!/_{3}\,[P_{flip}(\tau)+(1-P_{flip}(\tau)\gamma)] + {}^{2\beta_T}\!/_{3}\,\gamma$$

ii) $P_{T0}(\tau)$ : Probability for the final qubit state to be the $T_0$ after the manipulation.

- Initialize to $S$ ($\beta_S$), flip under the manipulation pulse ($P_{flip}(\tau)$), does not relax to the ground state (1-$\gamma$)
- Initialize to $T_0$ ($\beta_T/3$), does not flip under the manipulation pulse (1- $P_{flip}(\tau)$), does not relax to the ground state (1-$\gamma$)

$$\Rightarrow P_{T0}(\tau) = \beta_S P_{flip}(\tau)(1-\gamma) + {}^{\beta_T}\!/_{3}\,(1-P_{flip})(1-\gamma)$$

iii) $P_{T+}(\tau)$ ($P_{T-}(\tau)$) : Probability for the final qubit state to be the $T+$ ($T-$) after the manipulation.

- Initialize to T+ (T-) ($\beta_T/3$), does not relax to the ground state (1-$\gamma$)

$$\Rightarrow P_{T+}(\tau) = P_{T-}(\tau) = {}^{\beta_T}\!/_{3}\,(1-\gamma)$$

Combined with the tunneling detection infidelities, the probability for the tunneling event to be detected $P_D(\tau)$ can be calculated as, $P_D(\tau) = (P_{T0}(\tau) + P_{T+}(\tau) + P_{T-}(\tau))(1-E_T) + P_S(\tau)E_N + \alpha_S P_S(\tau)(1-E_T)$. We neglect the terms proportional to $E_T \cdot E_N$. By fitting the $P_D(\tau)$ to the measured Larmor oscillation (Fig. S4(b)), we extract the thermal tunneling error $\alpha_S \sim$ 0.6 %, and $\beta_T$ < 0.1 %. Note that the adaptive initialization scheme described above facilitates very low false initialization error $\beta_T$ and we expect the accurate measure of the $\beta_T$ should be possible with the self-consistent tomography schemes [11]. Also, large $E_{ST}/k_B T_e$ at the EST readout position provided by singlet-triplet splitting $E_{ST}$ on the order of 30 GHz [9] enables $\alpha_S$ < 1 %. Based on the error rates, we evaluate the singlet (triplet) measurement fidelity $F_S$ ($F_{T0}$) ~ 99.28 % (~ 98.53 %) yielding the total measurement fidelity about 99 %.

This corresponds to the quantum oscillation visibility of ~ 98 % consistent with the observation.

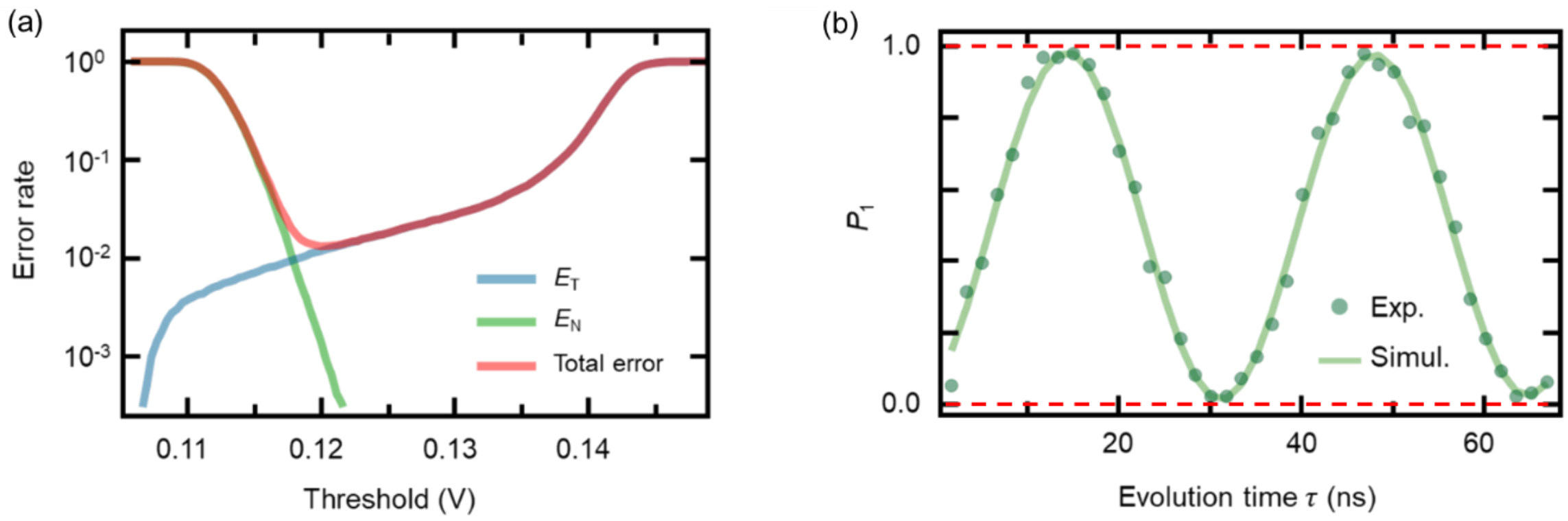

**Supplementary Figure S4. Quantum oscillation visibility analysis (a)** Tunneling (no-tunneling) detection infidelity shown in blue (green) curves. At the optimum threshold voltage, the error rate for the tunneling (no-tunneling) detection $E_T$ ($E_N$) ~ 1.4 % (0.7 %) is obtained. The red curve corresponds to the total error ($E_T + E_N$) as a function of the threshold voltage. **(b)** Experimental Larmor oscillation curve (green dot) and the numerical model (green curve) comprising the thermal tunneling, false initialization, and the relaxation errors. Fit to the model yield thermal tunneling error ($\alpha_S$) ~ 0.6 % with the false initialization error ($\beta_T$) < 0.1 %.

## S5. Randomized Benchmarking and Gate Set Tomography

*Randomized benchmarking* (RB and IRB): A single-qubit Clifford gate set is constructed using primitive gates I, X, Y, ±X/2, and ±Y/2, which are implemented by calibrated RF bursts. For concatenating RF bursts, we use an idle time of 16 ns. The elements of the Clifford gate set are randomly selected during the benchmarking. Each point in Fig. 4(c) is obtained by averaging 1000 single-shot measurements per sequence. The measurement data obtained from the standard randomized benchmarking (RB) is fitted to the exponentially decaying curve $P_1(m) = Ap_{\mathrm{avg}}^m + B$ where $m$ denotes the number of Clifford gates. The average gate fidelity $F_{\mathrm{avg}}$ is then determined by the depolarizing parameter $p_{\mathrm{avg}}$ as $\left(1+p_{\mathrm{avg}}\right)/2$ [12].

The gate fidelity of each primitive gate, on the other hand, is obtained with respect to the reference random Clifford gate sets using interleaved randomized benchmarking (IRB) protocol [12]. The measurement data from the interleaved randomized benchmarking is fitted to the same exponentially decaying curve $P_1(m) = Ap_{\text{gate}}^m + B$ to obtain the depolarizing parameter $p_{\text{gate}}$. The gate fidelity is then obtained as $\left(1 + p_{\text{gate}} / p_{\text{avg}}\right)/2$, where the effect of the reference RB is reflected as $1/p_{\text{avg}}$ [12].

*Gate set tomography* (GST): We use a single qubit gate set of {I, X/2, Y/2}, where the notation for each element is the same as those in the RB. Specifically, the length of all gates is fixed to a specific length, including the idle gate I. Compositing the elements in the gate set, we conducted the GST experiment with germs {I, X/2, Y/2, X/2∘Y/2, X/2∘X/2∘Y/2} and fiducials {null, X/2, Y/2, X/2∘X/2, X/2∘X/2∘X/2, Y/2∘Y/2∘Y/2} and the results are analyzed using the open-source python package, pyGSTi [13].

## Supplementary references